\documentclass[12pt]{article}
\usepackage{graphicx,epsfig}
\usepackage{amsmath,amssymb}
\usepackage{array}
\usepackage[utf8]{inputenc}
\usepackage{multicol}
\usepackage{cite}
\usepackage{color}
\usepackage{slashed}

\numberwithin{equation}{section}

\newcommand{\ord}[1]{\mathcal{O}\left({#1}\right)}
\newcommand{\eq}[1]{Eq.~(\ref{#1})}

\title{\Large\bf\boldmath
Probing the Type I Seesaw Mechanism with Displaced Vertices at the LHC
}
\date{}
\author{Alberto~M.~Gago$^{a}$\footnote{agago@pucp.edu.pe}, Pilar~Hern\'andez$^{b}$\footnote{m.pilar.hernandez@uv.es}, Joel~Jones-P\'erez$^{a,b}$\footnote{jones.j@pucp.edu.pe}, \\ Marta Losada$^{c}$\footnote{malosada@uan.edu.co}, Alexander Moreno Brice\~no$^{c}$\footnote{alexander.moreno@uan.edu.co}
\\[0.5 cm]
$^a${\em\normalsize Secci\'on F\'isica, Departamento de Ciencias, Pontificia Universidad Cat\'olica del Per\'u,} \\
{\em\normalsize  Apartado 1761, Lima, Peru} \\
$^b${\em\normalsize Instituto de F\'isica Corpuscular (IFIC), CSIC-Universitat de Val\`encia,} \\
{\em\normalsize Apartado de Correos 22085, E-46071 Valencia, Spain} \\
$^c${\em\normalsize Centro de Investigaciones, Universidad Antonio Nari\~no, Bogota, Colombia} 
}

\begin{document}

%\title{Probing the Type I Seesaw Mechanism with Displaced Vertices at the LHC}

%\author{Alberto~M.~Gago\inst{1}\thanks{agago@pucp.edu.pe} \and Pilar~Hern\'andez\inst{2}\thanks{m.pilar.hernandez@uv.es} \and Joel~Jones-P\'erez\inst{1,2}\thanks{jones.j@pucp.edu.pe} \and Marta Losada\inst{3}\thanks{malosada@uan.edu.co} \and Alexander Moreno Brice\~no\inst{3}\thanks{alexander.moreno@uan.edu.co}}

%\institute{Secci\'on F\'isica, Departamento de Ciencias, Pontificia Universidad Cat\'olica del Per\'u, Apartado 1761, Lima, Peru
%\and
%Instituto de F\'isica Corpuscular (IFIC), CSIC-Universitat de Val\`encia, \\ Apartado de Correos 22085, E-46071 Valencia, Spain
%\and
%Centro de Investigaciones en Ciencias Básicas y Aplicadas, Universidad Antonio Nari\~no, \\ Cra. 3 Este No. 47A-15, Bogotá D. C. 110231, Colombia}

%\date{Received: date / Revised version: date}

\maketitle

\begin{abstract}
The observation of Higgs decays into heavy neutrinos would be strong evidence for new physics associated to neutrino masses.  In this work we propose a search for such decays within the Type I seesaw model in the few-GeV mass range via displaced vertices. \\
Using 300 fb$^{-1}$ of integrated luminosity, at 13 TeV, we explore the region of parameter space where such decays are measurable. We show that, after imposing pseudorapidity cuts, there still exists a region where the number of events is larger than $\ord{10}$. We also find that conventional triggers can greatly limit the sensitivity of our signal, so we display several relevant kinematical distributions which might aid in the optimization of a dedicated trigger selection.
\end{abstract}

\section{Introduction}

The Type I Seesaw mechanism \cite{Minkowski:1977sc,GellMann:1980vs,Yanagida:1979as,Mohapatra:1979ia} is possibly the simplest extension of the Standard Model that can explain the smallness of neutrino masses. Even though most realizations of this mechanism invoke extra sterile neutrinos with Majorana masses too heavy to be probed, the possibility that these masses lie at the electroweak scale range is not excluded, and could actually be a more natural scenario.  Such a case generically requires small neutrino Yukawa couplings, of similar size as  those of the light charged leptons, however, this is not the only possibility. An approximate $U(1)_L$ lepton number symmetry can be imposed to protect the smallness of neutrino masses, allowing for larger Yukawa couplings and heavy masses at the electroweak scale \cite{Mohapatra:1986bd,Wyler:1982dd,Branco:1988ex}. These models imply new free parameters that cannot all be fixed by the light neutrino mass matrix, it is therefore of upmost importance to search for complementary tests. 

When the new heavy neutrinos are lighter than the Higgs, the latter can present novel decay channels, in particular, a decay into a light and a heavy neutrino~\cite{Pilaftsis:1991ug}. This would be followed by a subsequent decay of the heavy neutrino via a charged or neutral current interaction. In a number of recent references, the study of such Higgs decays at the LHC has been performed, focussing on the decay channels  $N\to\ell^+\ell^-\nu$ \cite{BhupalDev:2012zg} and $N\to\ell q q'$ \cite{Cely:2012bz}.

If the heavy neutrinos have masses of the order of a few GeV, the Higgs decay can lead to a noticeable displaced vertex, which is potentially a very powerful signal to look for~\cite{Gronau:1984ct}. This mass range is particularly interesting, because it might lead to successful baryogenesis \cite{Akhmedov:1998qx,Asaka:2005pn}.

Recently, the putative signal of a displaced vertex from heavy neutrinos produced in $W$ decays has been studied~\cite{Helo:2013esa,Izaguirre:2015pga}. In contrast to the latter work, in this paper we consider the signal of displaced vertices at the LHC, resulting from Higgs decays to heavy neutrinos. Such a measurement would allow us to directly probe the neutrino-Higgs coupling, giving a strong signal in favour of the Type-I seesaw model. This signal is obtained within the framework of a minimal 3+2 neutrino model with an approximate $U(1)_L$ symmetry\footnote{Similar analyses in more complicated models can be found, for instance, in~\cite{Cerdeno:2013oya,Strassler:2006ri} and Contribution 18 of~\cite{Brooijmans:2012yi}.}, and after imposing all existing constraints from neutrino masses, direct searches, neutrinoless double beta-decay and lepton flavour violating processes involving $\mu\leftrightarrow e$ transitions.

Although production of heavy neutrinos from Higgs decay is more limited statistically than that from $W$ decays, the two are sensitive to different combinations of parameters in the seesaw scenario, and are therefore complementary. The putative observation of both signals would be an unprecedented probe of the low-scale seesaw scenario. Conversely, the non-observation would impose stringent constraints that  may be essential to rule out an interesting range of seesaw scales.  

The paper is organised as follows. In Section~\ref{sec:model} we introduce the model and impose all existing constraints in a convenient parametrization. In Section~\ref{sec:higgsdecays}, we review the Higgs decays to heavy neutrinos and quantify the size of the corresponding branching ratios in the presently allowed parameter space. In Section~\ref{sec:results}, we consider Higgs production from gluon fusion at the LHC and study the displaced vertex signature. We illustrate the reach of an LHC search on the parameters space of the model, and discuss the impact of several kinematic cuts. In Section~\ref{sec:discussion} we conclude. 

Some useful formulae are presented in the appendices: in Appendix~\ref{app:0nubb} the contribution to neutrinoless double beta decay and in Appendix~\ref{app:hdecayformulae} the differential decay rate of the Higgs into a heavy and a light neutrino in the lab frame.

\section{Parametrization and Constraints}
\label{sec:model}

A minimal 3+2 neutrino model is characterized by the addition of two heavy sterile neutrinos. This is translated into a $5\times5$ neutrino mass matrix, which for the normal hierachy can be written in diagonal form as follows:
\begin{equation}
 \label{eq:nudef.nh}
 \mathcal{M}_\nu=U^*\,{\rm diag}(0,\,m_2,\,m_3,\,M_1,\,M_2)\,U^\dagger~,
\end{equation}
where $\mathcal{M}_\nu$ is in the basis where $Y_e$ is diagonal, and $m_i$ are mass ordered. The parametrization of~\cite{Donini:2012tt} decomposes $U$ into four blocks:
\begin{equation}
 U_{5\times5}=\left(\begin{array}{cc}
(U_{a\ell})_{3\times3} & (U_{ah})_{3\times2} \\
(U_{s\ell})_{2\times3} & (U_{sh})_{2\times2}
\end{array}\right)~.
\end{equation}
For the normal hierarchy, each block can be parametrized in the following way:
\begin{align}
 \label{eq:mixingmats}
 U_{a\ell} &= U_{\rm PMNS}\left(\begin{array}{cc}
 1 & 0 \\
 0 & H
\end{array}
\right)~, &
 U_{ah} &= i\,U_{\rm PMNS}\left(\begin{array}{c}
 0 \\
 H\,m_\ell^{1/2}R^\dagger M_h^{-1/2}
\end{array}\right)~, \nonumber \\
 U_{s\ell} &= i\left(\begin{array}{cc}
 0 & \bar H M_h^{-1/2}\,R\,m_\ell^{1/2}
\end{array}\right)~, &
 U_{sh} &= \bar H~,
\end{align}
where:
%\begin{eqnarray}
%\label{eq:hreal}
%H &=& \left(I+m_\ell^{1/2}\,R^\dagger\,M_h^{-1}\,R\,m_\ell^{1/2}\right)^{-1/2} \\
%\label{eq:hbarreal}
%\bar H &=& \left(I+M_h^{-1/2}\,R\,m_\ell\,R^\dagger\,M_h^{-1/2}\right)^{-1/2}~.
%\end{eqnarray}
\begin{align}
\label{eq:hreal}
H = \left(I+m_\ell^{1/2}\,R^\dagger\,M_h^{-1}\,R\,m_\ell^{1/2}\right)^{-1/2} & &
%\label{eq:hbarreal}
%\label{eq:hreal}
\bar H = \left(I+M_h^{-1/2}\,R\,m_\ell\,R^\dagger\,M_h^{-1/2}\right)^{-1/2}~.
\end{align}

In the previous equations, we have a unitary matrix $U_{\rm PMNS}$, which would correspond to the observed neutrino mixing matrix in the limit $H\to I$. The diagonal heavy (mostly sterile) neutrino mass matrix is denoted as $M_h={\rm diag}(M_1,\,M_2)$. The other two light (mostly active) massive neutrinos have a diagonal mass matrix denoted by $m_\ell={\rm diag}(m_2,\,m_3)={\rm diag}(\sqrt{\Delta m^2_{\rm sol}},\sqrt{\Delta m^2_{\rm atm}})$. Finally, we have a complex orthogonal matrix $R$~\cite{Casas:2001sr}, which is parametrized as:
\begin{equation}
 \label{eq:rmat}
 R=\left(\begin{array}{cc}
 \cos(\theta_{45}+i\gamma_{45}) & \sin(\theta_{45}+i\gamma_{45}) \\
 -\sin(\theta_{45}+i\gamma_{45}) & \cos(\theta_{45}+i\gamma_{45})
\end{array}\right)~.
\end{equation}
Thus, the only free parameters left in the neutrino mass matrix are the angles $\theta_{45}$, $\gamma_{45}$, the heavy neutrino masses $M_1$, $M_2$, and the two CP phases present in $U_{\rm PMNS}$. One can demonstrate that all of the mixing angles $\theta_{ij}$ can be restricted to the first quadrant.

If we want to express our results for the inverse hierarchy, we need to re-write Eq.~(\ref{eq:nudef.nh}) taking into account the appropriate ordering. This leads us to a different mixing matrix, $V$, such that:
\begin{equation}
 \mathcal{M}_\nu=V^*\,{\rm diag}(m_2,\,m_3,\,0,\,M_1,\,M_2)\,V^\dagger~.
\end{equation}
The reordering can be done through a permutation matrix acting on the active states. In blocks, we have:
\begin{align}
 \label{eq:mixingmats.ih}
 V_{a\ell} &= U_{\rm PMNS}\left(\begin{array}{cc}
 H & 0 \\
 0 & 1
\end{array}\right)~, &
 V_{ah} &= i\,U_{\rm PMNS}\left(\begin{array}{c}
 H\,m_\ell^{1/2}R^\dagger M_h^{-1/2} \\
 0
\end{array}\right)~, \nonumber \\
 V_{s\ell} &= i\left(\begin{array}{cc}
\bar H\,M_h^{-1/2}\,R\,m_\ell^{1/2} & 0
\end{array}\right)~, &
V_{sh} &= \bar H~.
\end{align}

Let us comment on the role of $\theta_{45}$ and $\gamma_{45}$. For large $|\gamma_{45}|\gtrsim2-3$, the hyperbolic sine and cosine in Eq.~(\ref{eq:rmat}) give essentially the same result (modulo a sign), such that $\theta_{45}$ behaves as an overall phase:
\begin{equation}
 R_{|\gamma_{45}|\gg1}=\left(\begin{array}{cc}
1 & \pm i \\
\mp i & 1
\end{array}\right)\cosh\gamma_{45}\,e^{\mp i\theta_{45}}~.
\end{equation}
Here, the $\pm$ refers to the sign of $\gamma_{45}$. As we can see, $\theta_{45}$ can be factorised out of the mixing, and plays no significant role within the phenomenology of the model. Thus, the relevant parameters in this model are the two heavy masses $M_1$, $M_2$, and the angle $\gamma_{45}$. Moreover, for fixed heavy masses, increasing $\gamma_{45}$ makes active-heavy mixing grow exponentially\footnote{Notice that the unitarity of the mixing matrix is kept under control by the $H$ and $\bar H$ matrices.}.

In order to understand better the active-heavy mixing in this limit, let us also assume $H\sim I$. In this case, for the normal hierarchy, we can write: 
\begin{eqnarray}
 U_{\ell4} \equiv (U_{ah})_{\ell 1} &=&  \pm Z^{\rm NH}_\ell\sqrt{\frac{m_3}{M_1}}\cosh\gamma_{45}\,e^{\mp i\theta_{45}}~, \\
 U_{\ell5} \equiv (U_{ah})_{\ell 2} &=&  i\,Z^{\rm NH}_\ell\sqrt{\frac{m_3}{M_2}}\cosh\gamma_{45}\,e^{\mp i\theta_{45}}~,
\end{eqnarray}
where:
\begin{equation}
 Z^{\rm NH}_\ell = (U_{\rm PMNS})_{\ell 3}\pm i\sqrt{\frac{m_2}{m_3}}(U_{\rm PMNS})_{\ell2}~.
\end{equation}
This structure is similar to that found in the literature (see for instance~\cite{Ibarra:2011xn}). In addition, up to corrections of $\ord{m_3/M_j}$, we can write the Dirac mass matrices, associated to the Yukawas, as:
\begin{eqnarray}
 (m_D)_{\ell 1} &=& \pm (Z^{\rm NH}_\ell)^*\sqrt{m_3 M_1}\cosh\gamma_{45}\,e^{\mp i\theta_{45}}~, \\
 (m_D)_{\ell 2} &=& -i (Z^{\rm NH}_\ell)^*\sqrt{m_3 M_2}\cosh\gamma_{45}\,e^{\mp i\theta_{45}}~.
\end{eqnarray}
Thus, in this limit the size of the Dirac masses is exponentially enhanced with respect to the naive expectations of Seesaw models, $m_D\sim\sqrt{m_i M_j}$. As an example, for values of $\gamma_{45}\sim7$, one would expect an enhancement of $\ord{10^3}$.

For the inverted hierarchy, the results are identical, but including:
\begin{equation}
 Z^{\rm IH}_\ell = (U_{\rm PMNS})_{\ell 2}\pm i\sqrt{\frac{m_2}{m_3}}(U_{\rm PMNS})_{\ell1}
\end{equation}

In the following, we shall not use any of the approximations above, and shall always take the exact form of $U_{ah}$ and $V_{ah}$. Also, for definiteness, we set all neutrino oscillation parameters to their best-fit points as in~\cite{Gonzalez-Garcia:2014bfa}, and all CPV phases to zero.

\subsection{Constraints}

\begin{figure}
\includegraphics[width=0.48\textwidth]{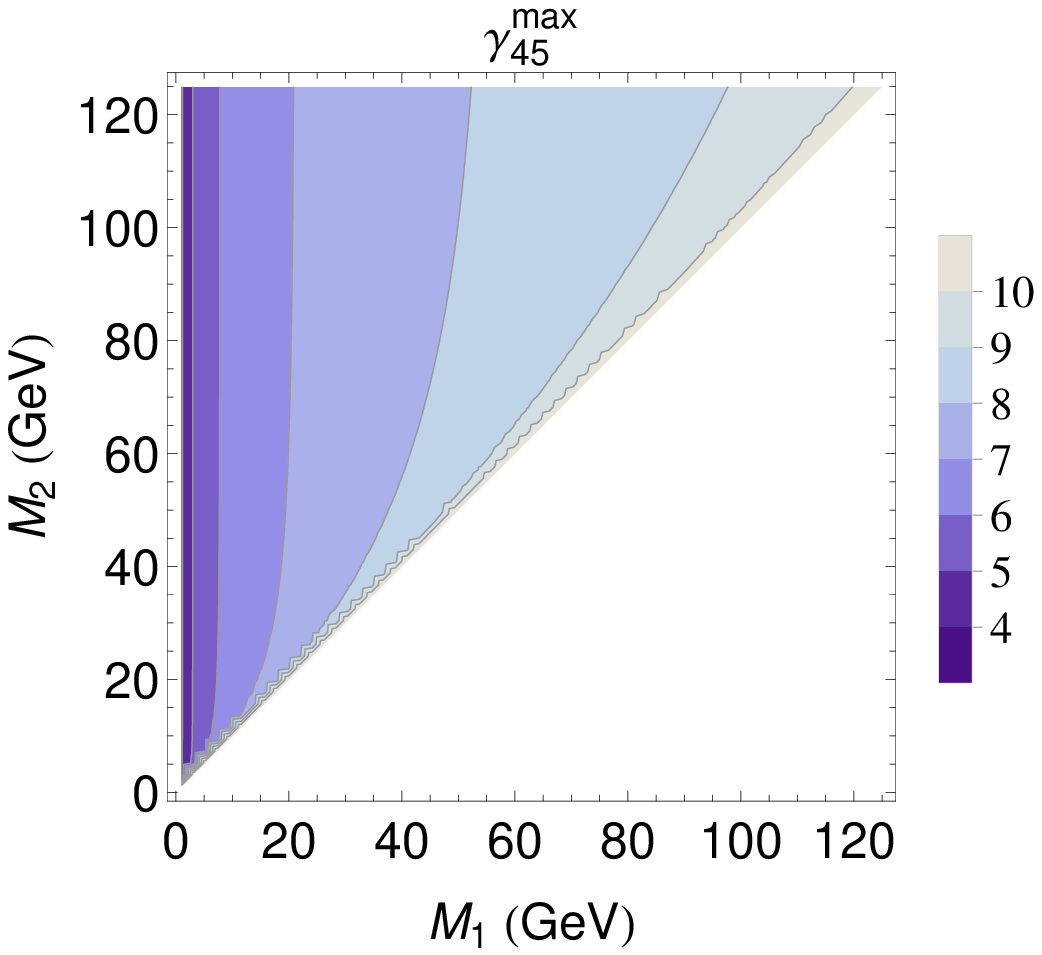}\quad
\includegraphics[width=0.4\textwidth]{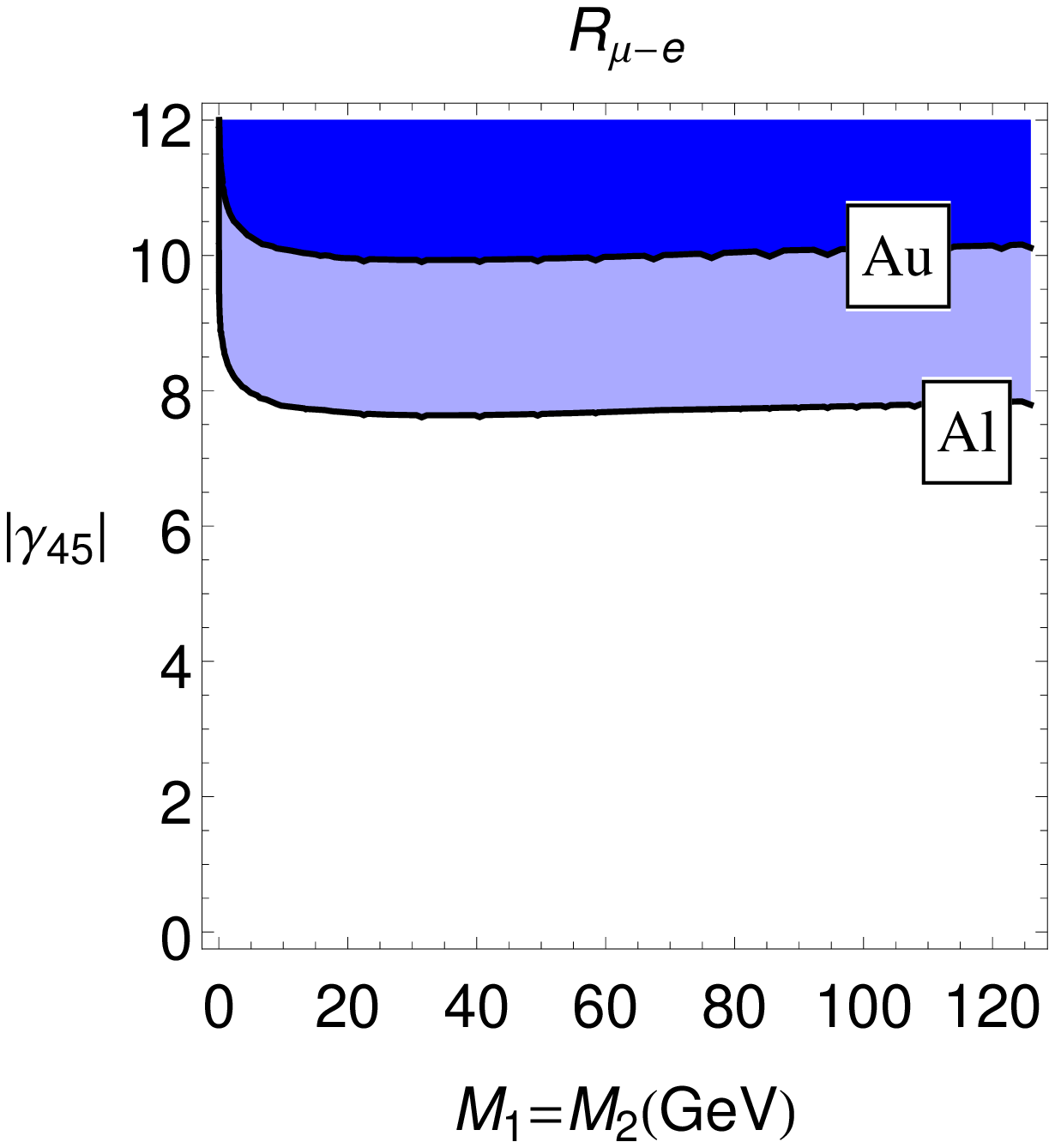}
\caption{Left: The contours show the maximum value of $\gamma_{45}$ allowed by the lack of observation of $0\nu\beta\beta$. Right: Bounds from $\mu-e$ conversion in nuclei, for the case of degenerate masses. The dark blue area is excluded by experiments with Au, while the light blue area can be probed in the future with Al experiments.} 
\label{fig:const1}
\end{figure}

There are three relevant constraints on the parameter space explored in this work. These come from neutrinoless double beta decay ($0\nu\beta\beta$), lepton flavour violation (LFV) and direct searches.

\subsubsection*{Neutrinoless Double Beta Decay}

Currently, the strongest constraints on $0\nu\beta\beta$ come from Germanium and Xenon experiments. On the Germanium front, the GERDA, HDM and IGEX experiments have combined their data, and determined a lower bound on the lifetime, $T^{0\nu}_{1/2}>3.0\times10^{25}$ yr. This corresponds to an effective mass $m_{\beta\beta}<0.2-0.4$~eV~\cite{Agostini:2013mzu}. On the other hand, for Xenon, KamLAND-Zen and EXO-200 have jointly imposed a lower bound of $T^{0\nu}_{1/2}>3.4\times10^{25}$ yr. This would correspond to $m_{\beta\beta}<0.12-0.25$~eV~\cite{Gando:2012zm}. A list of future $0\nu\beta\beta$ experiments can be found in~\cite{Schwingenheuer:2012zs,Gomez-Cadenas:2015twa}.

The non-observation of $0\nu\beta\beta$ can put very strong limits on the active-heavy mixing. To calculate this observable, we use the formulae derived in Appendix~\ref{app:0nubb}, based on the work in~\cite{Blennow:2010th}. The left panel of Figure~\ref{fig:const1} shows the maximum allowed value of $\gamma_{45}$ as a function of the two neutrino masses, for the normal hierachy, given the current bounds.

Notice that the bounds vanish for degenerate neutrinos, as expected from the second term in \eq{eq:0nubb}. It turns out that, for degenerate neutrinos, one can describe the neutrino mixing matrix with an inverse seesaw-like structure, making evident the existence of an approximate $U(1)_L$ symmetry. This symmetry constrains lepton-number violating processes from being too large. This was already discussed in~\cite{LopezPavon:2012zg}. Furthermore, in this limit, the light neutrino masses are protected from large loop corrections~\cite{AristizabalSierra:2011mn}. Thus, for the rest of this work, we shall consider the degenerate case, $M_1=M_2$.

\subsubsection*{Lepton Flavour Violation}

The most relevant processes constraining our parameter space are radiative LFV and $\mu-e$ conversion in nuclei.

Radiative LFV processes include $\mu\to e\gamma$ and $\tau\to\ell\gamma$ decays. For the former, the MEG experiment has placed an upper bound of ${\rm BR}(\mu\to e\gamma)<5.7\times10^{-13}$~\cite{Adam:2013mnn}, and the future upgrade expects to reach a value around $5\times10^{-14}$. For tau decays, both the Belle and BaBar experiments have constrained their branching ratios. The strongest ones are given by BaBar, of ${\rm BR}(\tau\to e\gamma)<3.3\times10^{-8}$ and ${\rm BR}(\tau\to \mu\gamma)<4.4\times10^{-8}$~\cite{Aubert:2009ag}. The future expected sensitivity for both channels at Belle II is of $\ord{10^{-9}}$.

Another important process is $\mu-e$ conversion in nuclei. The SINDRUM-II experiment has imposed limits on the conversion rate associated with Ti ($4.3\times10^{-12}$)~\cite{Dohmen:1993mp}, Au ($7\times10^{-13}$)~\cite{Bertl:2006up} and Pb ($4.6\times10^{-11}$)~\cite{Honecker:1996zf}. There exist several experiments which will attempt to probe lower values, such as Mu2e (Al, $\sim5.4\times10^{-17}$)~\cite{Mu2e}, COMET (Al, $\sim3\times10^{-17}$)~\cite{COMET}, and PRISM/PRIME (Ti, $\ord{10^{-18}}$)~\cite{Barlow:2011zza}.

To calculate all these processes, we use the formulae of~\cite{Alonso:2012ji}, and references within. We find that all observables give competitive constraints, but the most stringent, both now and in the future, comes from $\mu-e$ conversion. As none of these processes have yet been observed, an upper limit is imposed on $\gamma_{45}$, even for the degenerate case. This is shown on the right panel of Figure~\ref{fig:const1}.

\subsubsection*{Direct searches}

\begin{figure}
\includegraphics[width=0.31\textwidth]{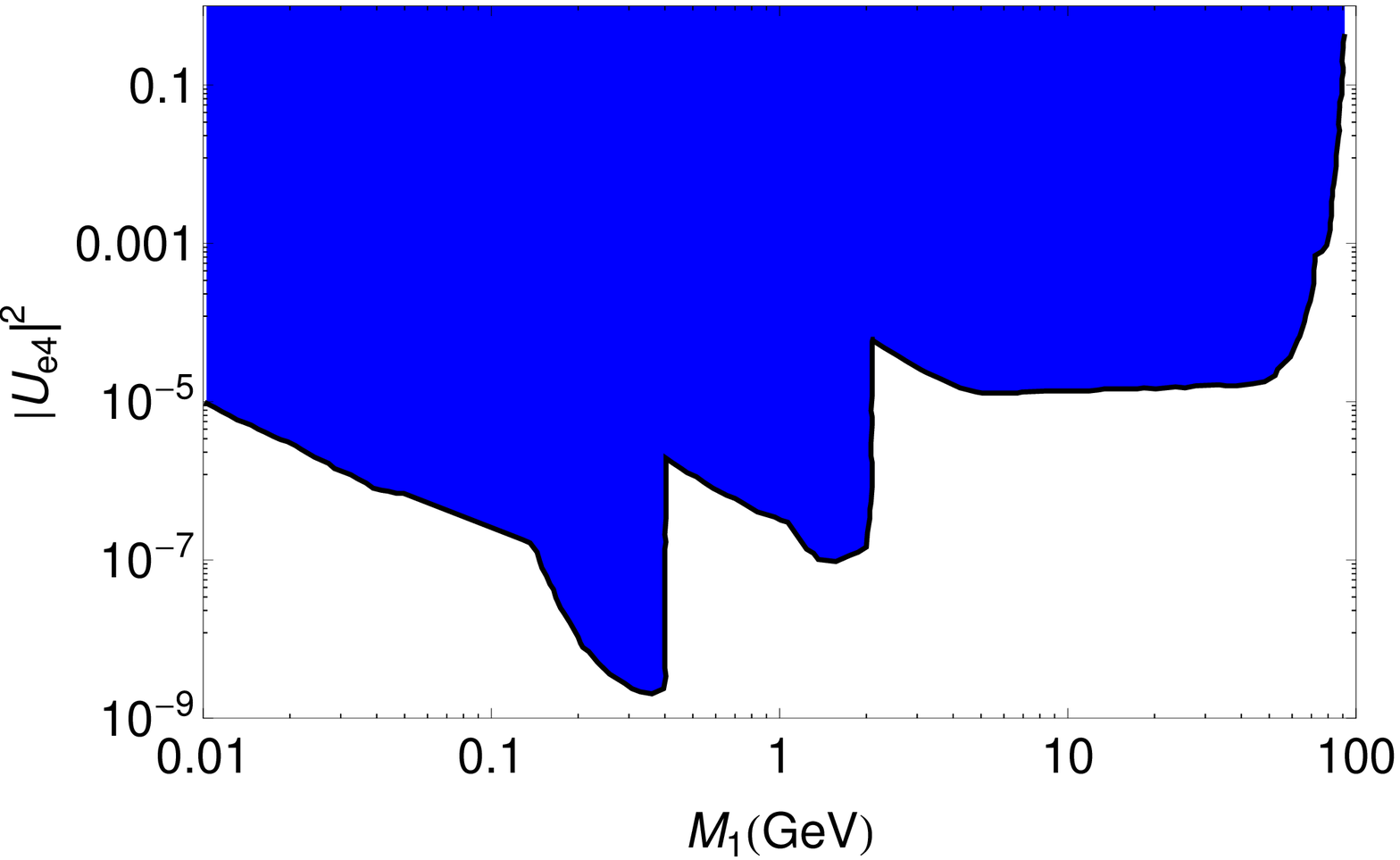}\quad
\includegraphics[width=0.31\textwidth]{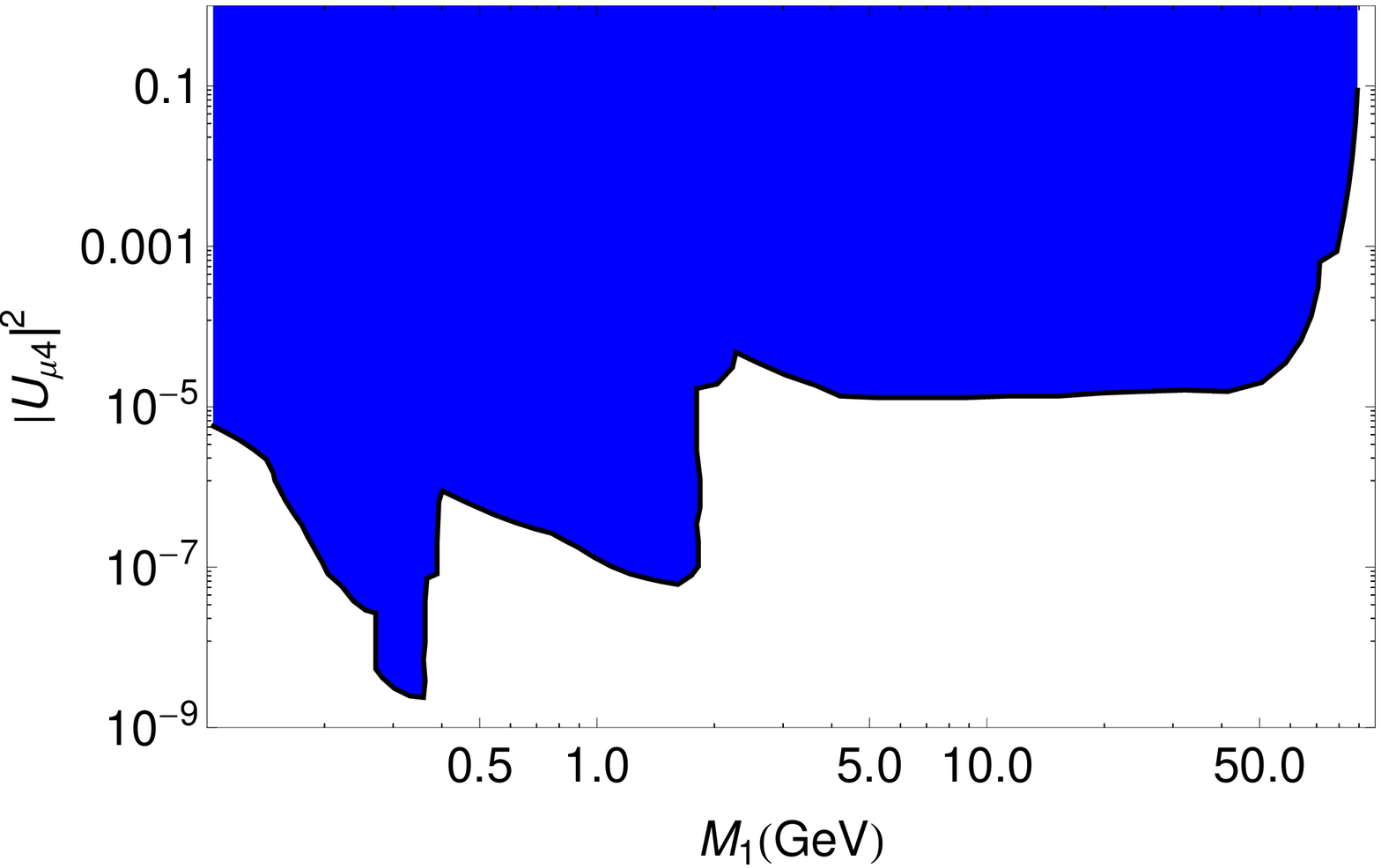}\quad
\includegraphics[width=0.31\textwidth]{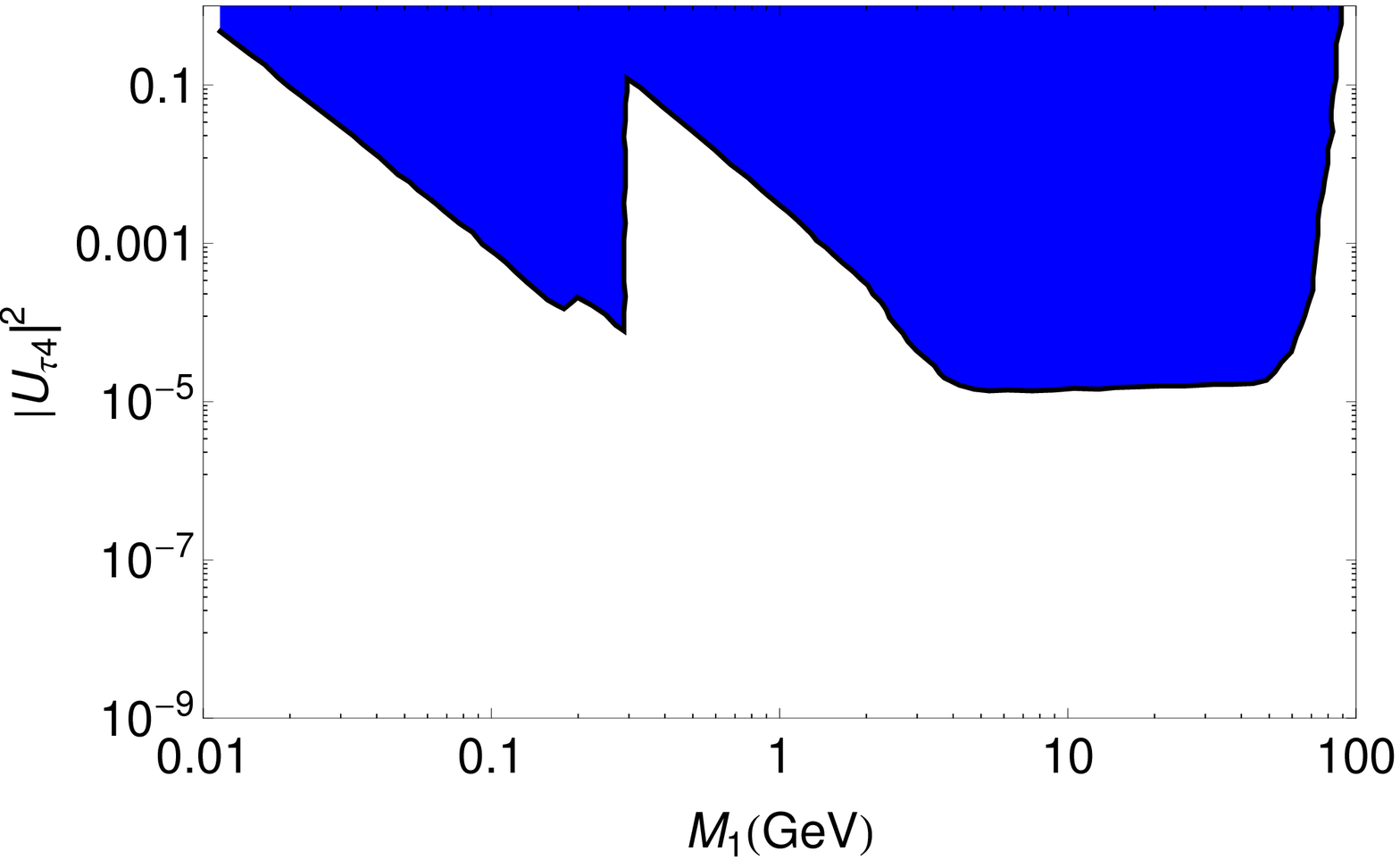}
\caption{Constraints placed on the heavy neutrino parameter space, due to direct searches. We show constraints for $|U_{e4}|^2$, $|U_{\mu4}|^2$ and $|U_{\tau4}|^2$ on the left, center and right panels, respectively.}
\label{fig:directsearch}
\end{figure}

Finally, we also need to apply direct search bounds. Many experiments have tried to produce, and detect, these heavy neutrinos. Again, the lack of observation puts constraints on active-heavy mixing. Providing a faithful interpretation of each result on the 3+2 model is beyond the scope of this work. Thus, we shall take the bounds as reported in~\cite{Atre:2009rg}, shown in Figure~\ref{fig:directsearch}, and apply them directly to our framework.

The most important direct search constraints for this work are those of DELPHI~\cite{Abreu:1996pa}. One must note that, although all three bounds by DELPHI seem competitive, the importance of one or another shall depend on the predictions for the mixing within the model. For instance, for the normal hierarchy, one finds that generally $|U_{e4}|^2$ is smaller than $|U_{\mu4}|^2$ and $|U_{\tau4}|^2$ by an order of magnitude, meaning that the latter two bounds shall be more stringent.

\section{Higgs Decays into Heavy Neutrinos}
\label{sec:higgsdecays}

As mentioned previously, observing Higgs decay into neutrinos would be a strong signal in favour of the seesaw model. The Higgs partial decay width into two neutrinos was initially calculated in~\cite{Pilaftsis:1991ug}, and can be written as:
\begin{equation}
\label{eq:Hf1f2}
\Gamma(h\to \nu_i \nu_j) = \frac{\omega}{8\pi m_h}\lambda^{1/2}(m_h^2,\,m_{\nu_i}^2,\,m_{\nu_j}^2) 
\left[S\left(1-\frac{(m_{\nu_i}+m_{\nu_j})^2}{m_h^2}\right)+P\left(1-\frac{(m_{\nu_i}-m_{\nu_j})^2}{m_h^2}\right)\right]
\end{equation}
where $m_h$ is the Higgs mass, and the scalar and pseudoscalar couplings are:
\begin{align}
S&=\frac{g^2}{4m^2_W}\left((m_{\nu_i}+m_{\nu_j})\,\Re e[C_{ij}]\right)^2 ~, &
P&=\frac{g^2}{4m^2_W}\left((m_{\nu_j}-m_{\nu_i})\,\Im m[C_{ij}]\right)^2~,
\end{align}
with $C_{ij}=\sum_{k=1}^3U^*_{ki}U_{kj}$. Moreover, $\lambda(a,b,c)=a^2+b^2+c^2-2ab-2bc-2ac$ is a kinematic function, and $\omega=1/n!$ for $n$ identical final states. We find that the largest branching ratio happens for the decay into one light and one heavy neutrino:
\begin{equation}
\label{eq:1light1heavy1}
 \Gamma(h\to n_iN_j)=\frac{g^2}{32\pi}\frac{M_j^2}{m_W^2}m_h\left(1-y_j^2\right)^2\left|C_{ij}\right|^2~,
\end{equation}
where $y_j=M_j/m_h$ and, for the normal hierarchy:
\begin{equation}
 C_{ij}=i\left(\begin{array}{c}
0 \\ H^2 m_\ell^{1/2}R^\dagger M_h^{-1/2}
\end{array}\right)~.
\end{equation}
For the inverted hierarchy, one shifts the $(2-3)$ rows to the $(1-2)$ rows. Here we see the very important fact that the PMNS matrix does not appear in the partial width. In particular, this means that our results shall not depend on the unknown Majorana nor Dirac CP phases. This is not the case for heavy neutrino searches involving the $W$ boson, where the active-heavy mixing plays a central role due to the presence of charged leptons. %{\color{red} {\it Removed by Marta:} For instance, as can be seen in \eq{eq:ueh}, the PMNS appears non-trivially within $U_{eh}$, implying that CP phases can induce significant variations.}
This shows that, within this framework, the measurement of both $h\to n_i N_j$ and $W\to \ell N_j$ decays would be complementary, with the possibility to access the value of the CP phases.

\begin{figure}
 \centering
 \parbox{0.47\textwidth}{
 \resizebox{0.45\textwidth}{!}{\includegraphics{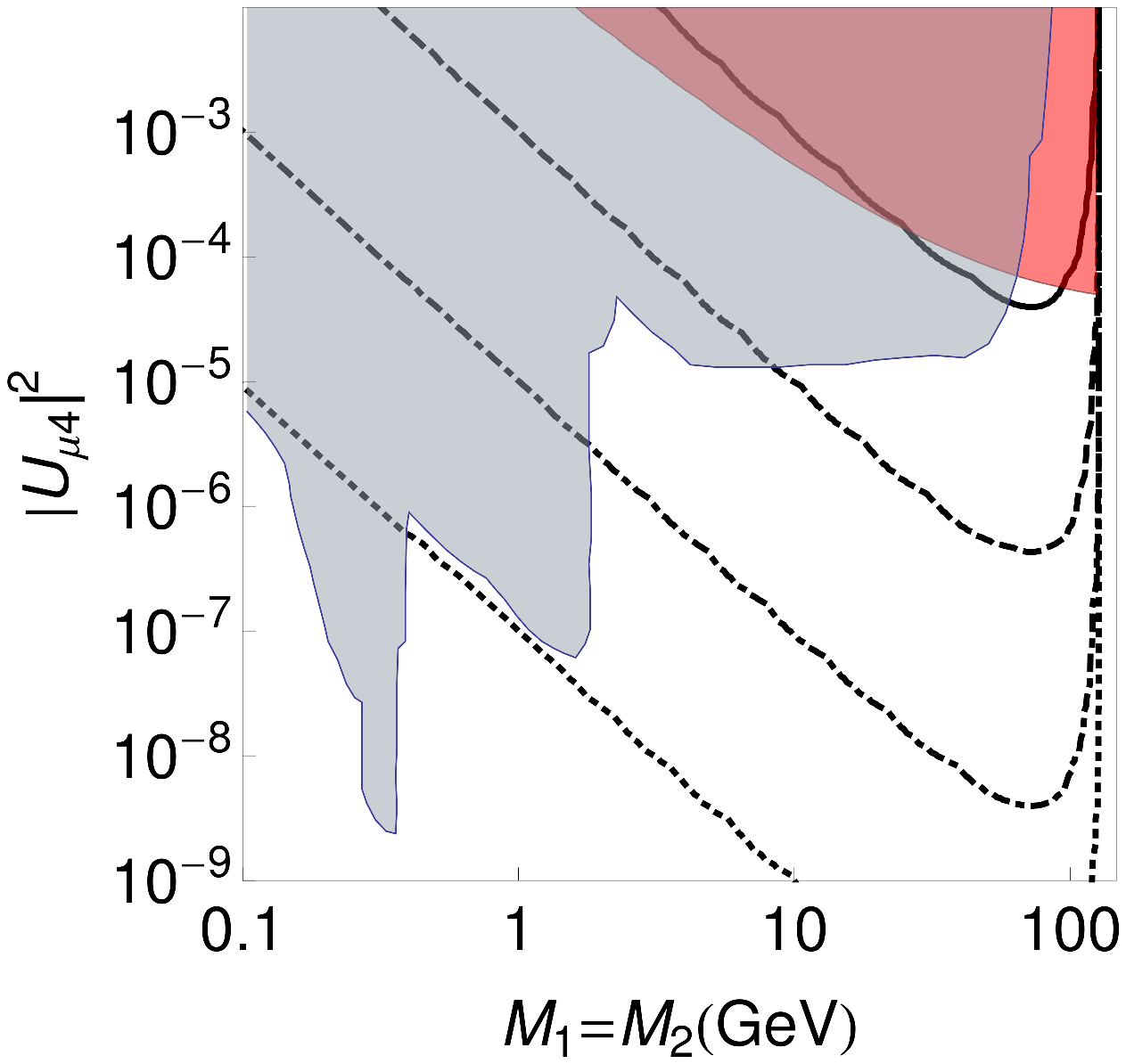}}
\caption{Branching ratio for $h\to n N$. LFV and direct search constraints are shown in red and blue, respectively. Contours indicate branching ratios of $10^{-2}$, $10^{-4}$, $10^{-6}$ and $10^{-8}$, from inner to outer curve.} 
\label{fig:hnunuBR}}\qquad
\parbox{0.47\textwidth}{
\resizebox{0.43\textwidth}{!}{\includegraphics{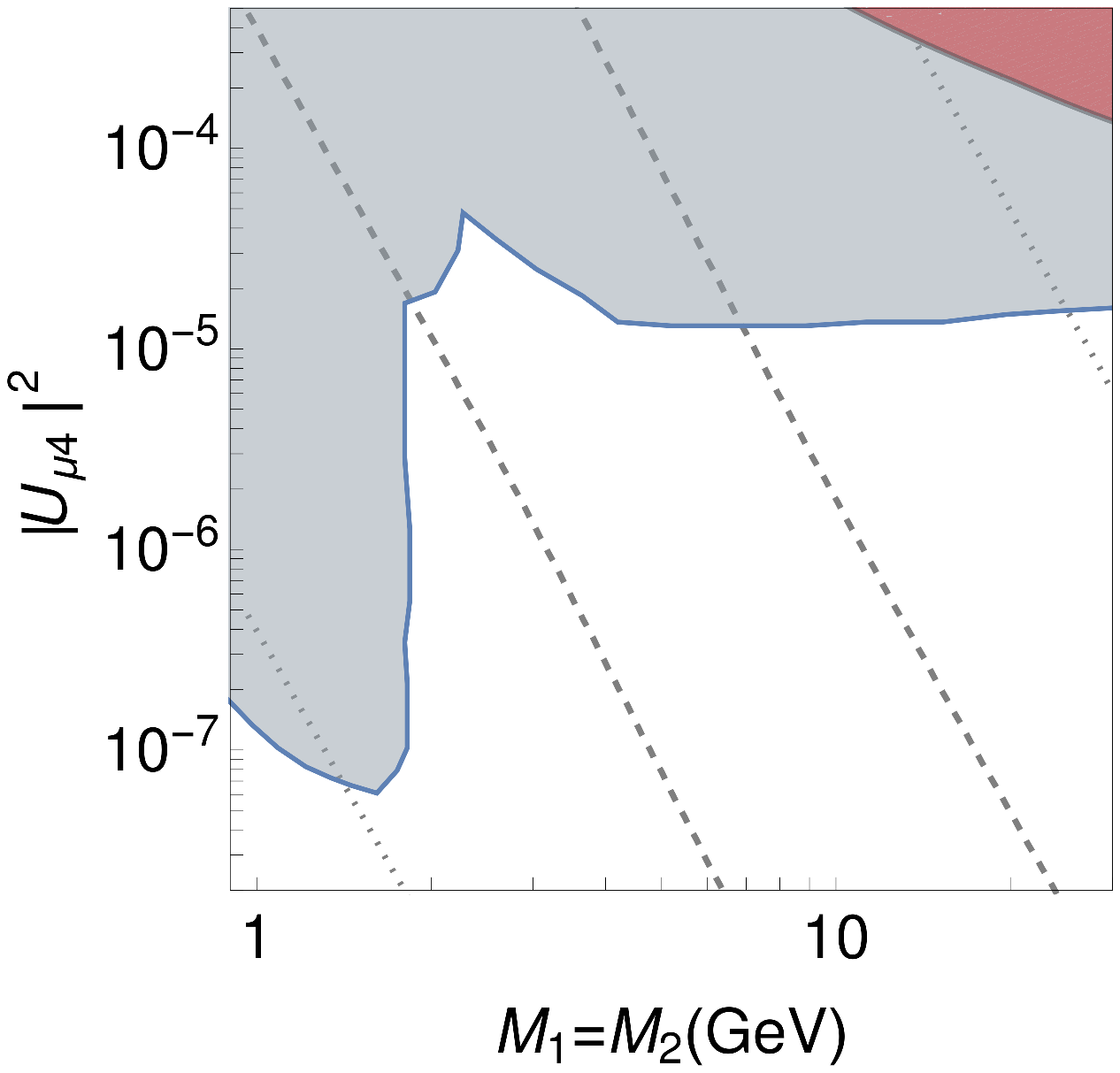}}
\caption{Decay length, $\tau_Nc$. Constraints are shown as in Figure~\ref{fig:hnunuBR}. The region between dashed lines has $1\,{\rm mm}\leq\tau_Nc\leq10^3\,{\rm mm}$, dotted lines indicate $10^{-3}\,{\rm mm}\leq\tau_Nc\leq10^{6}\,{\rm mm}$. The decay length decreases from left to right.} 
\label{fig:declen}}
\end{figure}

We plot the $h\to n_i N_j$ branching ratio in Figure~\ref{fig:hnunuBR} for the normal hierarchy, along with constraints from LFV and direct searches. We find that branching ratios as large as $0.01$ are generally ruled out by the former constraints. Moreover, if we want to work with branching ratios large enough to provide a signal at the LHC, the heavy masses cannot have values under a few GeV.

\section{Displaced Vertices from Higgs Decays}
\label{sec:results}

The heavy neutrinos are not stable, and eventually decay through charged and neutral current interactions. The decay channels, width and lifetime can be found, for instance, in~\cite{Atre:2009rg,Kovalenko:2009td}. If the neutrino transverse decay length lies between 1~mm and 1~m, a displaced vertex signal could be recorded at ATLAS and CMS~\cite{Aad:2012zx,CMS:2014hka,CMS:2014wda,Aad:2015uaa,Aad:2015rba,Khachatryan:2014mea}.

We now consider the possibility of observing such a signal, as a product of Higgs boson decays. The first step is to identify the region of interest, that is, one where the decay length is acceptable and the Higgs branching ratio is not too small. The transverse decay length $\ell_{N_T}$ is related to the heavy neutrino lifetime $\tau_N$ through:
\begin{equation}
 \label{eq:decaylength}
\ell_{N_T}=\frac{|\vec{p}_{N_T}|}{M_j}\tau_N c~,
\end{equation}
where both the heavy neutrino mass $M_j$ and transverse momentum $\vec{p}_{N_T}$ are measured in GeV. Thus, for a given mass, the requirement of having a visible $\ell_{N_T}$ puts constraints on $|\vec{p}_{N_T}|\tau_N$.

In order to get an approximate idea of the region of interest, we plot in Figure~\ref{fig:declen} the decay length $\tau_Nc$. The shape of the curves can be understood by realizing that $\tau_Nc\propto M_j^5\left|U_{\ell4}\right|^2$ and then taking the logarithm. One needs to be aware that these curves are given only to roughly illustrate the region where displaced vertices might be visible. The parameter which ultimately defines the region is the transverse decay length $\ell_{N_T}$, which depends on the transverse momentum with which the neutrino is generated.

The $\tau_Nc$ lines are not parallel to the contour lines for the $h\to n_i N_j$ branching ratio. This means that at some point the region of interest shall have a too small probability for $h\to n_i N_j$ decay. Thus, we find that we require heavy neutrino masses between 2-20 GeV to be able to probe a displaced signature without significantly reducing the expected number of events.

We now estimate the number of displaced vertices in this region. We take Higgs production through gluon fusion, $gg\to h$, followed by the decay $h\to n_i N_j$. If we do not consider specific final states after heavy neutrino decay, nor any kinematical cuts, the event rate with a measurable displaced vertex at the LHC is:
\begin{eqnarray}
 \label{eq:nodisp}
 N&=&\mathcal{L}\int d\left|\vec{p}_{h_T}\right|dy_h\, d\left|\vec{p}_{N_T}\right|d\phi_N \nonumber \\ 
 &&\times\frac{d^2\sigma(gg\to h)}{d\left|\vec{p}_{h_T}\right|dy_h}
 \frac{\gamma_h}{\Gamma_h}\frac{d^2\Gamma(h\to n_i N_j)}{d\left|\vec{p}_{N_T}\right|d\phi_N}
 \Theta_H\left(\ell_{N_T}{\rm (mm)}-1\right)\Theta_H\left(10^3-\ell_{N_T}{\rm (mm)}\right)
\end{eqnarray}
Here, $\vec{p}_{h_T}$ and $y_h$ are the Higgs transverse momentum and rapidity. The azimuthal angle between $\vec{p}_{h_T}$ and $\vec{p}_{N_T}$ is denoted by $\phi_N$. In addition, $\Gamma_h$ is the Higgs width, while $\gamma_h$ is the corresponding relativistic factor, $\gamma_h=E_h/m_h$. The luminosity $\mathcal{L}$ is taken equal to $300$~fb$^{-1}$ for the second run of the LHC. The two Heaviside $\Theta_H$ functions make sure the decay length lies within the detection capability. The integration is performed with the Vegas subroutine of the {\tt CUBA} library~\cite{Hahn:2004fe}.

The integrand contains two differential distributions. The first one corresponds to $gg\to h$ production, as a function of the  Higgs transverse momentum and rapidity. This is obtained through the codes {\tt SusHi}~\cite{Harlander:2005rq,Harlander:2012pb} and {\tt MoRe-SusHi}~\cite{Mantler:2012bj,Harlander:2014uea}. The second is the Higgs differential decay width, in the lab frame, for decay into one light and one heavy neutrino. This is shown in Appendix~\ref{app:hdecayformulae}. Note that we are using cylindrical coordinates. This is done in order to directly constrain the transverse decay length through the integration limits for $|\vec{p}_{N_T}|$.

\begin{figure}
\begin{center}
\resizebox{0.45\textwidth}{!}{%
\includegraphics[width=0.45\textwidth]{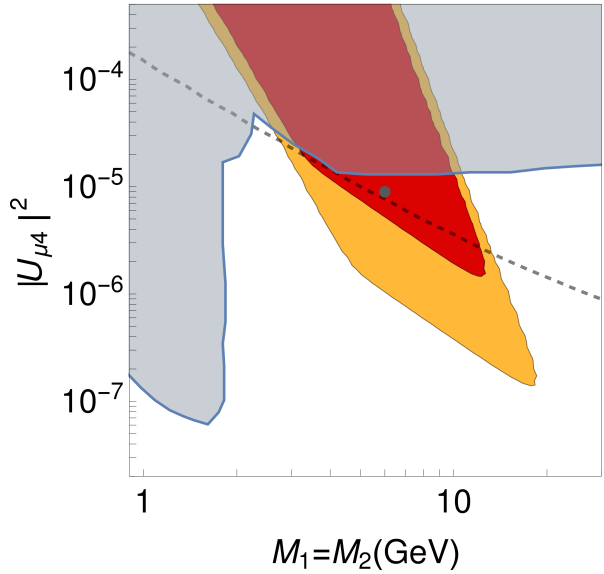}}
\end{center}
\caption{Region sensitive to events with a displaced vertex. The region in red would have more than 250 events with a displaced vertex, for an integrated luminosity of $300\,$fb$^{-1}$, at 13 TeV. The region in orange would have more than 50 events. The blue region is ruled out by direct searches, and the dashed line indicates the reach of $\mu-e$ conversion experiments using Al nuclei. The dot represents a benchmark point.} 
\label{fig:nevents}
\end{figure}

The region of parameter space leading to events with a visible displaced vertex is shown in Figure~\ref{fig:nevents}, for the normal hierarchy. We show the regions excluded by direct searches in blue, and the reach of future LFV experiments by the dashed curve. The region in red would have more than 250 events with a displaced vertex, using $300\,$fb$^{-1}$ of integrated luminosity at 13 TeV. The region in orange would have 50 events for the same luminosity.

As mentioned previously, the left and right boundaries are determined by the experimental requirements on the decay length. For smaller masses, the decay length is too large, and the heavy neutrino escapes the detector. For larger masses, the decay length is too small, and the detector resolution is incapable of discriminating the displaced vertex from the interaction point. This constraint is imposed by introducing Eq.~(\ref{eq:decaylength}), which depends on the heavy neutrino mass and mixing, within the Heaviside functions in Eq.~(\ref{eq:nodisp}). This result also depends on the Higgs branching ratio being large enough, which determines the lower boundary of the region.

Thus, we see that regions with heavy neutrino masses between 2-20 GeV favour displaced vertex events at the LHC, for values of $|U_{\mu4}|^2$ between $\ord{10^{-7}}$ and $\ord{10^{-5}}$. For future comparison, we establish a benchmark point for the normal hierarchy, with $M_1=M_2=6\,$GeV and $\gamma_{45}=8$. This point leads to 428 events with a detectable displaced vertex, and is displayed as a dot in Figure~\ref{fig:nevents}.

\subsection{Signatures from Heavy Neutrino Decays}

The previous section allowed us to calculate the number of events with a displaced vertex happening due to Higgs decays. However, these events are not necessarily observable. The heavy neutrino eventually decays into other final state particles, which need to be detected.

We obtain the differential decay rate for heavy neutrinos, and convolute it in Eq.~(\ref{eq:nodisp}). Since the heavy neutrino is lighter than the $W$ boson, two-body decays are not allowed. Therefore, it will decay through a three-body process. Given its relatively large branching ratio, we focus on $N_j\to\mu qq'$ decay, where the momenta of the final states is labelled by $p_1$, $p_2$ and $p_3$, respectively. To calculate the differential decay rate, we follow~\cite{Greiner:1993qp}. The procedure is carried out in two frames. First, part of the integration is done in the frame where the spatial component of $p_N-p_3$ vanishes. On this frame, the momentum components shall be denoted with a tilde (i.e.\ $\tilde\phi_1$). The rest of the integration is then performed in the frame where the heavy neutrino is at rest. Momentum components in this frame shall be denoted with a hat (i.e\, $\hat\theta_3$). Finally, we relate these variables with the appropriate ones in the lab frame, as this is where the experimental cuts are placed. These shall be denoted with a ``lab'' superscript (i.e.\ $\vec{p}_1^{\,{\rm lab}}$). 

The observed number of events is given by:
\begin{equation}
 \label{eq:diffdec2}
 N=\mathcal{L}\int d\Omega\frac{d^2\sigma(gg\to h)}{d\left|\vec{p}_{h_T}\right|dy_h}
 \frac{\gamma_h}{\Gamma_h}\frac{d^2\Gamma(h\to n_i N_j)}{d\left|\vec{p}_{N_T}\right|d\phi_N}
 \frac{\gamma_N}{\Gamma_N}\frac{d\Gamma(N_j\to\mu^- q \bar q')}{ds\,d(\cos\hat\theta_3)d(\cos\tilde\theta_1)d\hat\phi_3d\tilde\phi_1} F_{\rm cuts}~,
\end{equation}
where the neutrino differential decay width can be found in~\cite{Greiner:1993qp}, and:
\begin{equation}
 d\Omega = d\left|\vec{p}_{h_T}\right|dy_h\, d\left|\vec{p}_{N_T}\right|d\phi_Nds\,d(\cos\hat\theta_3)d(\cos\tilde\theta_1)d\hat\phi_3d\tilde\phi_1~. 
\end{equation}
The integration is carried out over four new angular variables, as well as $s=(p_N-p_3)^2/M_j^2$.

Furthermore, $F_{\rm cuts}$ is a function describing all experimental cuts. For instance, if the experiment was to impose cuts on the transverse momenta and pseudorapidities of each final state, we would have:
\begin{eqnarray}
 F_{\rm cuts} &=& \Theta_H\left(\ell_{N_T}{\rm (mm)}-1\right)\Theta_H\left(10^3-\ell_{N_T}{\rm (mm)}\right) \\
 &&\quad\times\Pi_i\Theta_H(|\vec{p}_{i_T}|^{\rm lab}-|\vec{p}_{i_T}|^{\rm min})\,\Theta_H(\eta_i^{\rm max}-\eta_i^{\rm lab})~.
\end{eqnarray}

\begin{figure}
\begin{center}
\includegraphics[width=0.45\textwidth]{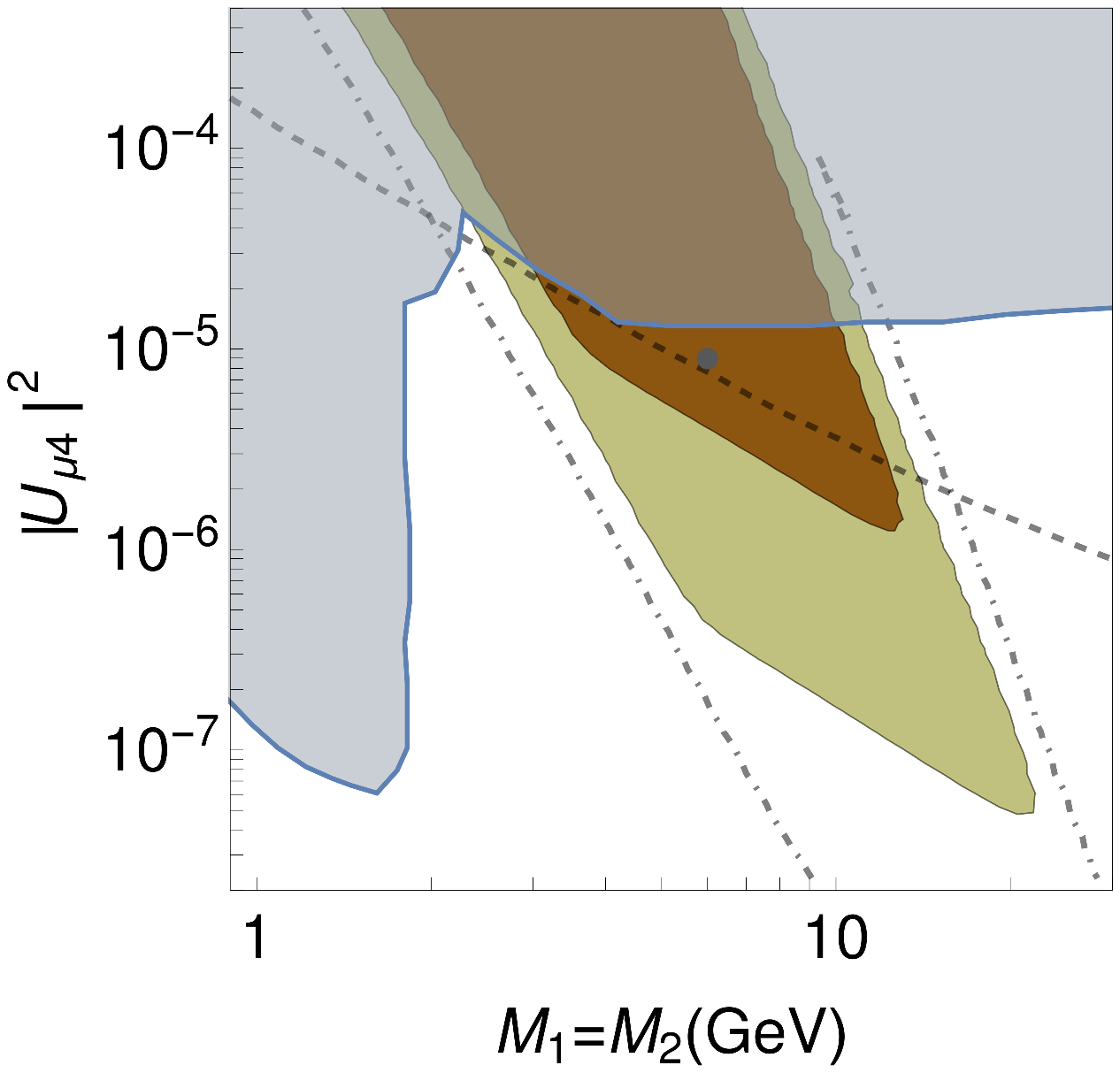} \quad
\includegraphics[width=0.45\textwidth]{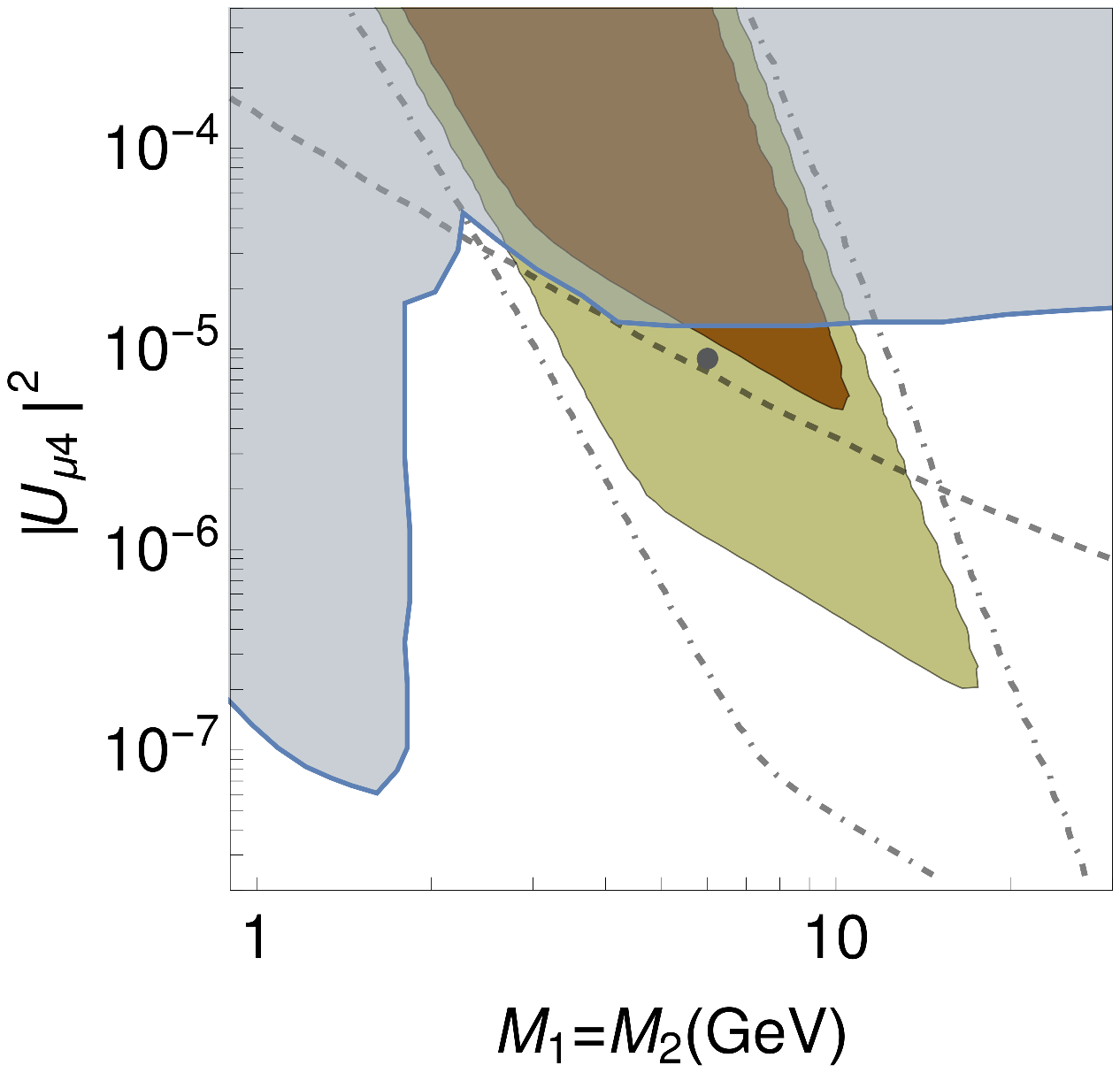}
\end{center}
\caption{Similar to Figure~\ref{fig:nevents}, number of events in the $h\to nN\to n\mu qq'$ channel. The region in brown (green) would have more than 100 (10) events with a displaced vertex. The dash-dotted line indicates more than one event. On the left we show the region if no cuts are applied to the final states, on the right we only apply pseudorapidity cuts on all final states.} 
\label{fig:nevents.final}
\end{figure}
To get a better understanding of the sensitivity of this signal, we plot on the left panel of Figure~\ref{fig:nevents.final} the region with visible displaced vertices, assuming no cuts on the final state. Our result essentially shows the same information as Figure~\ref{fig:nevents}, weighted by the $N\to\mu qq'$ branching ratio for each point. This scales the number of events by a factor $1/4$ - $1/5$. The right panel of Figure~\ref{fig:nevents.final} shows the same region, but including also conventional pseudorapidity cuts, that is, $|\eta_\mu|<2.4$ for the muon, and $|\eta_i|<2.5$ for every other particle. We find that, although the overall shape of the region remains unchanged, the number of events is affected by the cut.

As an example, we report the results for our benchmark point. In Figure~\ref{fig:nevents.final}, we have a total of 110 events on the $h\to n_i N_j\to\mu q q'$ channel with no cuts (left panel), which is further reduced to 78 events once all pseudorapidity constraints are applied (right panel). This is to be compared to the 428 events we expected from $h\to n_i N_j$ decays (Figure~\ref{fig:nevents}).

\subsection{Impact due to Kinematical Cuts}

For the purpose of giving a perspective of a future experimental search, we discuss the impact of several kinematic cuts on our analysis.

In the following, we plot the ratio of surviving events for each cut, imposing at the same time the displaced vertex and pseudorapidity constraints previously discussed. In order to understand the impact of the heavy neutrino mass, we show results for $M_j=3,\,15\,$GeV, which are limiting values of our signal region. We find that the results are not strongly influenced by the heavy neutrino mass. For each cut, we also compare the exact number of events for our benchmark point.

\begin{figure}
\begin{center}
\resizebox{0.45\textwidth}{!}{%
\includegraphics{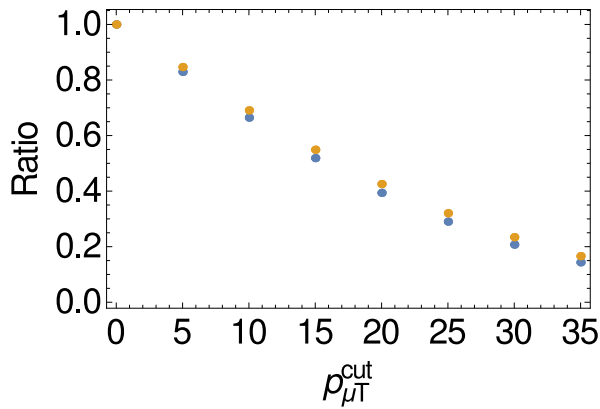}}
\end{center}
\caption{Ratio between the number of events with and without a cut on the muon transverse momentum, for the $h\to n_i N_j\to n_i \mu q q'$ channel. The blue (orange) points represent $M_i=3\,$GeV ($15\,$GeV).} 
\label{fig:ptCuts}
\end{figure}
The first constraint we study is a cut on the transverse momentum of the muon, $p^{\rm cut}_{\mu_T}$. This is shown in Figure~\ref{fig:ptCuts}. We find that typical cuts between 20 and 30 GeV would reduce the number of events to a total between $40\%$ and $20\%$. As an example, the benchmark point shows 32 (17) events after imposing a 20 (30) GeV cut. %We also show the impact of a 20 GeV cut on the whole parameter space on Figure~\ref{fig:nevents:ptcut}, in order to better understand the implications of such a loss of events.
%\begin{figure}
%\resizebox{0.4\textwidth}{!}{%
%\includegraphics[width=0.45\textwidth]{nevents-moresushi.mu4.vegas.eps}}
%\caption{Similar to Figure~\ref{fig:nevents.final}, but with an additional 20 GeV cut of $|\vec{p}_{\mu_T}|$. The region in brown (green) would have more than 100 (10) events with a displaced vertex. The dash-dotted line indicates more than one event.} 
%\label{fig:nevents:ptcut}
%\end{figure}
Since in this analysis we are not including detector effects, such as efficiency, it is clear that we need to relax the stringency of $p^{\rm cut}_{\mu_T}$ if we want to significantly improve the sensitivity with respect to that from DELPHI. We consider that, as these muons are not produced at the interaction point, a dedicated trigger with a smaller cut on the transverse momentum is more appropriate.

\begin{figure}
 \centering
 \parbox{0.47\textwidth}{
 \resizebox{0.45\textwidth}{!}{\includegraphics{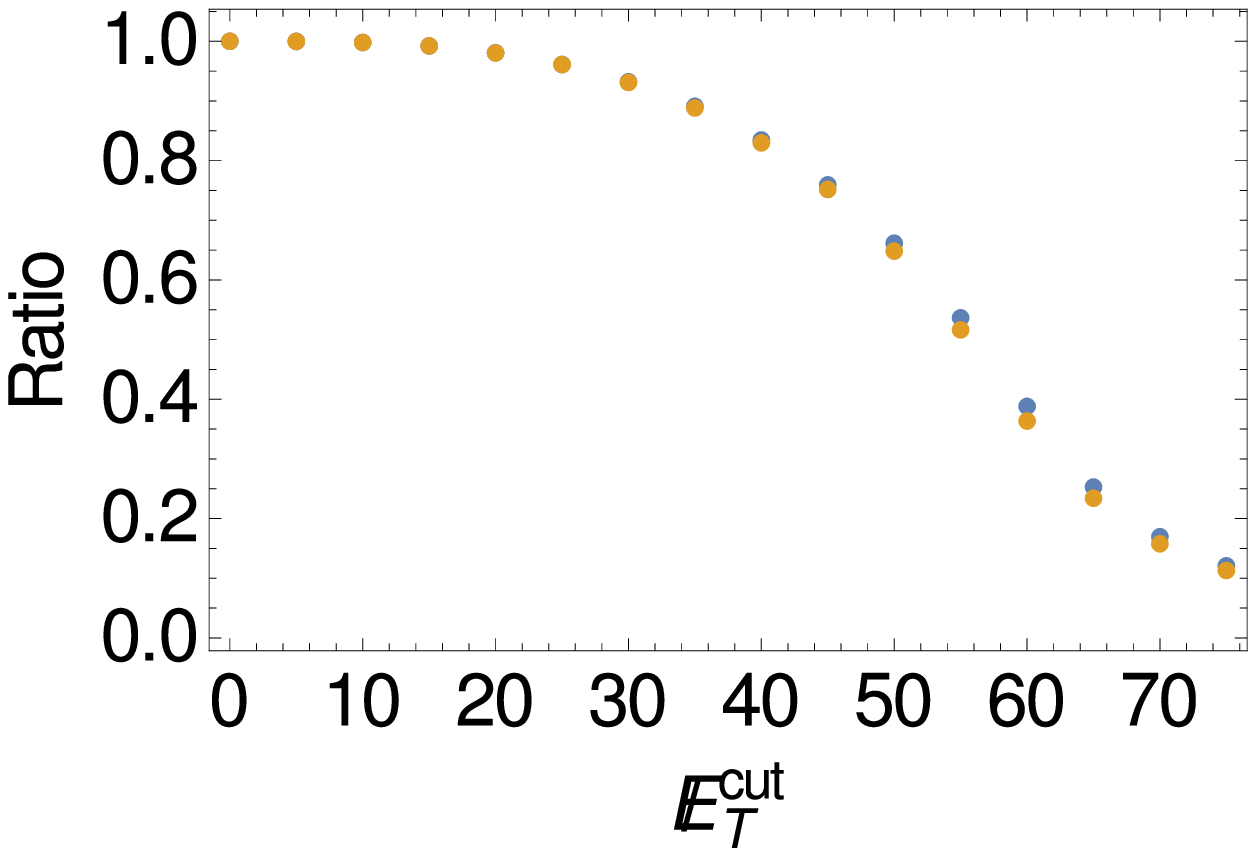}}
\caption{As Figure~\ref{fig:ptCuts}, but applying a cut on missing transverse energy.} 
\label{fig:etCuts}}\qquad
\parbox{0.47\textwidth}{
\resizebox{0.45\textwidth}{!}{\includegraphics{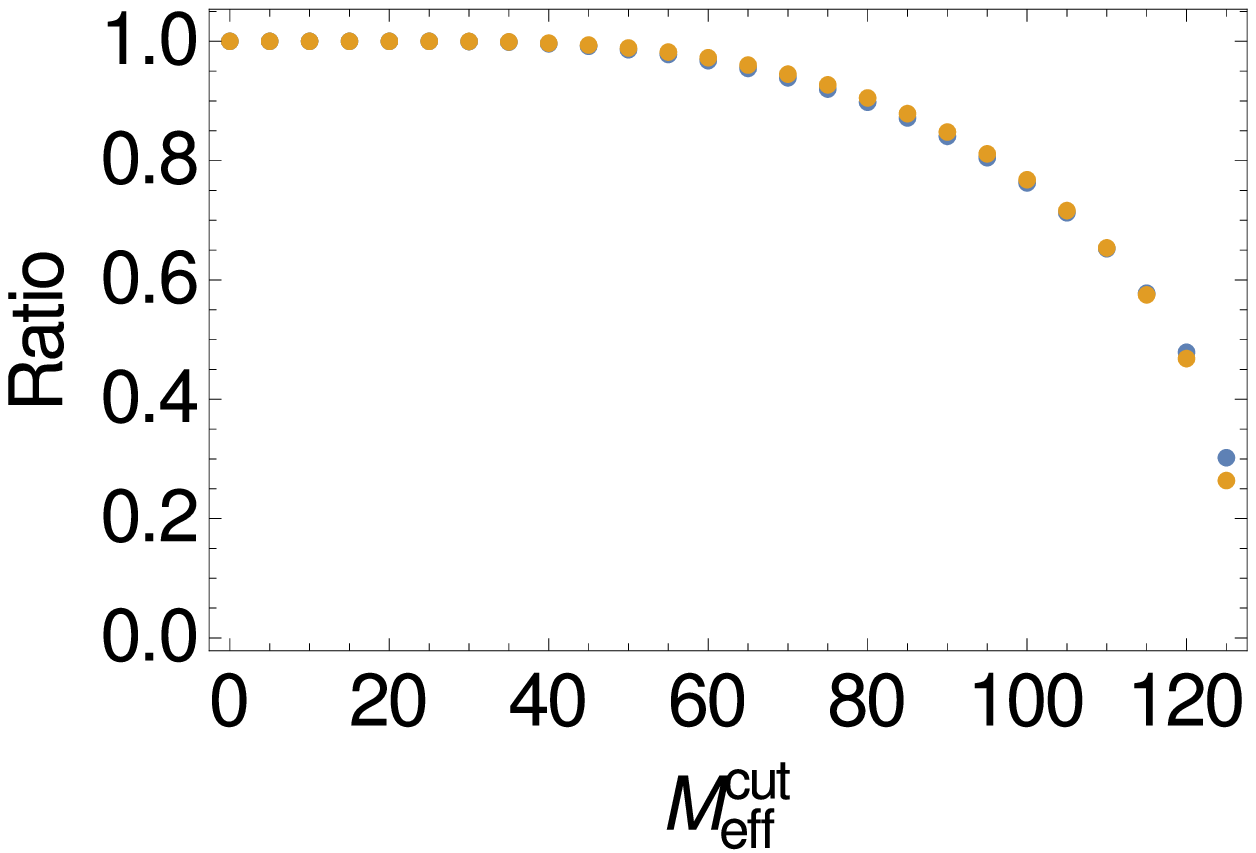}}
\caption{As Figure~\ref{fig:ptCuts}, but applying a cut on $M_{\rm eff}$.} 
\label{fig:MeffCuts}}
\end{figure}

Another cut of interest is that on missing transverse energy, $\slashed{E}_T=|\vec{p}_{\nu_T}|$, shown on Figure~\ref{fig:etCuts}. We observe that one can impose $\slashed{E}_T$ cuts up to 40 GeV without reducing the number of events below $80\%$. However, at this point the ratio drops, and one find that for cuts above 70 GeV the ratio is again under $20\%$. For our benchmark point, imposing displaced vertex, pseudorapidity and $\slashed{E}_T$ constraints, we find 75 (15) events for a $\slashed{E}_T$ cut of 40 (70) GeV.
%\begin{figure}
%\begin{center}\resizebox{0.45\textwidth}{!}{%
%\includegraphics{eTcut.eps}}
%\end{center}
%\caption{As Figure~\ref{fig:ptCuts}, but applying a cut on missing transverse energy.} 
%\label{fig:etCuts}
%\end{figure}

%\begin{figure}
%\begin{center}
%\resizebox{0.45\textwidth}{!}{%
%\includegraphics{Meffcut.eps}}
% \end{center}
%\caption{As Figure~\ref{fig:ptCuts}, but applying a cut on $M_{\rm eff}$.} 
%\label{fig:MeffCuts}
%\end{figure}

Finally, in Figure~\ref{fig:MeffCuts}, we show the impact of a cut on $M_{\rm eff}=|\vec{p}_{\mu_T}|+|\vec{p}_{q_T}|+|\vec{p}_{q'_T}|+\slashed{E}_T$. As expected, we find an endpoint for $M_{\rm eff}=m_h$. In this case it is possible to impose cuts as large as 95 GeV without reducing the number of events under $80\%$. Again, on our benchmark point, cutting on $M_{\rm eff}=95\,$GeV decreases the number of events to 63. 

\section{Discussion}
\label{sec:discussion}

In this work, we study the possible observation of Higgs decays involving heavy neutrinos, by means of a search for displaced vertices.

This study is done in the context of the minimal 3+2 neutrino model, which is based on a Type-I Seesaw with two heavy sterile neutrinos. After imposing all constraints on the parameter space, we find that the model can be described in terms of two additional parameters, apart from the light neutrino masses and mixings. The two new parameters are a degenerate mass for the two heavy neutrinos, and the enhancement parameter in the Casas-Ibarra $R$ matrix, $\gamma_{45}$.

We then calculate the partial width for Higgs decay into any two neutrinos. We find that the $h\to n_i N_j$ channel has the largest branching ratio, and concentrate on the description of a displaced vertex signal. This signal is particularly relevant for degenerate heavy neutrino masses of the order of a few GeV.

It is important to stress that this prediction depends on the neutrino masses and on $\gamma_{45}$, with no dependence on the neutrino mixing angles nor phases of the PMNS matrix. Such decays can therefore provide direct information on the new parameters of the model.

For the LHC Run 2, there exist allowed regions of parameter space where the number of Higgs decays with a displaced vertex could be as large as $\ord{100}$, before any other kinematical cut. It is important to note that the observation of the displaced vertex relies strongly on the decay channel of the heavy neutrinos and on the detection efficiency. As an example, we have included the branching ratio due to $N\to\mu qq'$ decay, and imposed pseudorapidity cuts on the final states. Both considerations reduce the number of events down to $\ord{20\%}$ from the original number, still leaving a large enough amount to be observed.

In order to perform a more realistic assessment of the signal strength, we have considered additional kinematical cuts. For instance, in the $N\to\mu qq'$ channel, we find that a 30 GeV cut on the muon transverse momentum, (typical of a level 1 trigger), the number of observable events is reduced to $4\%$ of the initial number. This low efficiency is due to the low value of the momenta of final states, which in turn is a consequence of the low mass of the heavy neutrinos. Therefore, as this does not include further potential losses from detector reconstruction inefficiencies, we conclude that one cannot rely on conventional cuts to properly observe this channel. In order to avoid this situation, we present two alternative kinematical cuts with much better efficiency. Such cuts are based on missing transverse energy, $\slashed{E}_T$, and effective mass, $M_{\rm eff}$. We believe this could be helpful in designing a dedicated trigger, and point out that such a trigger could be useful more generally to search for weakly interacting light particles.

The measurement of Higgs decays to heavy neutrinos would constitute a powerful test of the mechanism of neutrino mass generation.  This process can provide complementary information to the one that can be measured via the  dominant production mechanism, $W\to\ell N_j$ decays.

\section{Acknowledgements}
We would like to thank Stefan Liebler and Marius Wiesemann for support on SusHi and MoRe-SusHi. We would also like to thank Matt King for useful discussions on searches with displaced vertices.

A.G.~and J.J.P.~acknowledge funding by the Direcci\'on de Gesti\'on de la Investigaci\'on at PUCP, through grant DGI-2015-3-0026. J.J.P.~would like to thank CERN for its hospitality, and acknowledges partial support from the grants Generalitat Valenciana VALi+d, Spanish MINECO FPA 2011-23596 and the Generalitat Valenciana PROMETEO - 2008/004. P.H.~was supported by grants FPA2011-29678, PROMETEO/2009/116, CUP (CSD2008 - 00037), ITN INVISIBLES (Marie Curie Actions, PITN-GA-2011-289442). M.L.~and A.M.B.~were supported by grants of the VCTI-UAN/20131044 and the ITN INVISIBLES (Marie Curie Actions, PITN-GA-2011-289442). M.L.~would like to acknowledge hospitality at IFIC, Valencia and PUCP, Lima. A.M.B.~acknowledges the hospitality at IFIC, Valencia.

\appendix

\section{Neutrinoless Double Beta Decay}
\label{app:0nubb}

Following~\cite{Blennow:2010th}, we find that the $0\nu\beta\beta$ amplitude $A$ is proportional to:
\begin{equation}
 \label{eq:neutrinoless}
 A\propto \sum_{\ell_i=1}^3{m_{\ell i}\,U_{e\ell_i}^2\,\mathcal{M}^{0\nu\beta\beta}(m_{\ell_i})} + \sum_{i=1}^2{M_i\,U_{eh_i}^2\,\mathcal{M}^{0\nu\beta\beta}(M_i)}~,
\end{equation}
where $\mathcal{M}^{0\nu\beta\beta}$ is the nuclear matrix element. Furthermore, for heavy neutrinos with mass larger than 100 MeV, one has:
\begin{eqnarray}
 \mathcal{M}^{0\nu\beta\beta}(m_{\ell_i}) &\to& \mathcal{M}^{0\nu\beta\beta}(0)~, \\
 \mathcal{M}^{0\nu\beta\beta}(M_i) &\to& 0~.
\end{eqnarray}
It is common in the literature to define:
\begin{equation}
 m_{\beta\beta}=\sum_{\ell_i=1}^3{m_{\ell i}\,U_{e\ell_i}^2}=-\sum_{i=1}^2{M_i\,U_{eh_i}^2}~,
\end{equation}
where the last equality is guaranteed by the seesaw mechanism, at tree level. Then, for non-degenerate masses, we can understand the heavy neutrino contribution by writing:

\begin{eqnarray}
\sum_{i=1}^2{M_i\,U_{eh_i}^2}\,\mathcal{M}^{0\nu\beta\beta}(M_i) &=& 
\left(\sum_{i=1}^2{M_i\,U_{eh_i}^2}\right)\,\mathcal{M}^{0\nu\beta\beta}(M_2)
+\left(\mathcal{M}^{0\nu\beta\beta}(M_1)-\mathcal{M}^{0\nu\beta\beta}(M_2)\right)M_1\,U_{eh_1}^2 \nonumber \\
&=& -m_{\beta\beta}\,\mathcal{M}^{0\nu\beta\beta}(M_2)+\Delta\mathcal{M}^{0\nu\beta\beta}M_1\,U_{eh_1}^2 
\end{eqnarray}

The first term on the right is proportional to the contribution from the light neutrinos, but strongly suppressed by the matrix element involving the heaviest neutrino. Thus, for heavy neutrino masses larger than 100 MeV, the second term would provide the dominant contribution. 

In this model, the matrix element $U_{eh_i}$ can be written:
\begin{equation}
 \label{eq:ueh}
 U_{eh_i} = i\left[S^e_{\rm sol}\,R^\dagger_{1i}+S^e_{\rm atm}\,R^\dagger_{2i}\right] M_i^{-1/2}~.
\end{equation}

For the normal hierarchy we have:
\begin{align}
& \left(S^e_{\rm sol}\right)_{\rm NH} =
 \left[(U_{\rm PMNS})_{12}H_{11}+(U_{\rm PMNS})_{13}H_{21}\right](\Delta m^2_{\rm sol})^{1/4} \\
&
 \left(S^e_{\rm atm}\right)_{\rm NH} = 
 \left[(U_{\rm PMNS})_{12}H_{12}+(U_{\rm PMNS})_{13}H_{22}\right](\Delta m^2_{\rm atm})^{1/4}
\end{align}
while for the inverted hierarchy:
\begin{align}
& \left(S^e_{\rm sol}\right)_{\rm IH} = \left[(U_{\rm PMNS})_{11}H_{11}+(U_{\rm PMNS})_{12}H_{21}\right](\Delta m^2_{\rm sol})^{1/4} \\
& \left(S^e_{\rm atm}\right)_{\rm IH} =  \left[(U_{\rm PMNS})_{11}H_{12}+(U_{\rm PMNS})_{12}H_{22}\right](\Delta m^2_{\rm atm})^{1/4}
\end{align}

Then, the whole amplitude is proportional to:
\begin{eqnarray}
\label{eq:0nubb}
A &\propto&  m_{\beta\beta}\left(\mathcal{M}^{0\nu\beta\beta}(0)-\mathcal{M}^{0\nu\beta\beta}(M_2)\right) \nonumber \\
  &&-\frac{1}{4}\Delta\mathcal{M}^{0\nu\beta\beta}\left[(S^e_{\rm sol}+S^e_{\rm atm})\,e^{i(\theta_{45}-i\gamma_{45})}
+(S^e_{\rm sol}-S^e_{\rm atm})\,e^{-i(\theta_{45}-i\gamma_{45})}\right]^2
\end{eqnarray}
Here, we see that the amplitude can be exponentially enhanced by $\gamma_{45}$, negating the suppression from the matrix elements. For very large $\gamma_{45}$, the only way to control this enhancement is by having degenerate heavy neutrino masses.

\section{Higgs Decays into Heavy Neutrinos}
\label{app:hdecayformulae}

For completeness, we report the differential Higgs decay width, in cylindrical coordinates, on a boosted frame. We take a vanishing light neutrino mass, and take the heavy neutrino mass equal to $M_j$:

\begin{multline}
 \label{eq:diffhdec}
 \frac{d^2\Gamma(h\to n_i N_j)}{d\left|\vec{p}_{N_T}\right|d\phi_N}=
 \frac{1}{8\pi^2}\frac{|\vec{p}_{N_T}|}{\sqrt{m_h^2+|\vec{p}_{h_T}|^2}}\left[(S+P)\frac{m_h^2}{2}\left(1-\frac{M_j^2}{m_h^2}\right)\right] \\
 \times\left[\left|\frac{1}{p_{N_Z}(E_1+E_2)-p_{h_Z}E_1}\right|_{p_{N_Z}=p^+_{Z}}+\left|\frac{1}{p_{N_Z}(E_1+E_2)- p_{h_Z} E_1 }\right|_{p_{N_Z}=p^-_{Z}}\right]~.
\end{multline}
Here, the energy of the outgoing neutrinos is:
%\begin{equation}
% E_1 = \sqrt{M_j^2+|\vec{p}_{N_T}|^2+p_{N_Z}^2}
% \end{equation}
% \begin{multline}
% E_2 = \big[|\vec{p}_{N_T}|^2+p_{N_Z}^2+|\vec{p}_{h_T}|^2+p_{h_Z}^2 \\
% -2|\vec{p}_{h_T}||\vec{p}_{N_T}|\cos\phi_N -2p_{h_Z}p_{N_Z}\big]^{1/2}~.
%\end{multline}
\begin{eqnarray}
E_1 &=& \sqrt{M_j^2+|\vec{p}_{N_T}|^2+p_{N_Z}^2} \\
E_2 &=& \big[|\vec{p}_{N_T}|^2+p_{N_Z}^2+|\vec{p}_{h_T}|^2+p_{h_Z}^2
 -2|\vec{p}_{h_T}||\vec{p}_{N_T}|\cos\phi_N -2p_{h_Z}p_{N_Z}\big]^{1/2}~.
\end{eqnarray}
The Higgs momentum in the direction of the beam axis is defined in terms of the transverse momentum and rapidity:
\begin{equation}
 p_{h_Z}=\sqrt{m_h^2+|\vec{p}_{h_T}|^2}\sinh y_h
\end{equation}

Similarly, the variable $p_{N_Z}$ is the heavy neutrino momentum on the direction of the beam axis. It is fixed by momentum conservation, and has the following allowed values:

\begin{multline}
 p_Z^\pm = \frac{1}{2(m_h^2+|\vec{p}_{h_T}|^2)}
 \bigg{\{} \left(m_h^2+M_j^2+2|\vec{p}_{N_T}||\vec{p}_{h_T}|\cos\phi_N\right)p_{h_Z} \\
 \pm\bigg{[}(m_h^2+|\vec{p}_{h_T}|^2+p_{h_Z}^2) 
 \bigg{(}(m_h^2-M_j^2)^2-4(m_h^2|\vec{p}_{N_T}|^2+M_j^2|\vec{p}_{h_T}|^2) \\
 -4|\vec{p}_{N_T}|^2|\vec{p}_{h_T}|^2\sin^2\phi_N
 +4(m_h^2+M_j^2)|\vec{p}_{N_T}||\vec{p}_{h_T}|\cos\phi_N\bigg{)}\bigg{]}^{1/2}\bigg{\}}
\end{multline}

Demanding $p_Z^\pm$ to have real values puts constraints on $|\vec{p}_{N_T}|$ and $\phi_N$. We find that, if:
\begin{equation}
 |\vec{p}_{h_T}|\leqslant\frac{(m_h^2-M_j^2)}{2M_j} \quad\Rightarrow\quad 
 \left\{\begin{array}{c}
  0\leqslant|\vec{p}_{N_T}|\leqslant p_T^+ \\
  -\pi\leqslant\phi_N\leqslant\pi
 \end{array}\right.
\end{equation}
Alternatively, if:
\begin{equation}
 \frac{(m_h^2-M_j^2)}{2M_j}<|\vec{p}_{h_T}| \quad\Rightarrow\quad
 \left\{
 \begin{array}{c}
 % -\pi/2\leqslant\phi_N\leqslant\pi/2 \\
   p_T^-\leqslant|\vec{p}_{N_T}|\leqslant p_T^+\\
   \sin^2\phi_N\leqslant\frac{(m_h^2-M_j^2)^2}{4M_j^2|\vec{p}_{h_T}|^2}
 \end{array}\right.
\end{equation}

The values $p_T^\pm$ are defined as:

\begin{multline}
 p_T^\pm = \frac{1}{2(m_h^2+|\vec{p}_{h_T}|^2\sin^2\phi_N)}\bigg{\{}(m_h^2+M_j^2)|\vec{p}_{h_T}|\cos\phi_N \\
 \pm \bigg{[}(m_h^2+|\vec{p}_{h_T}|^2)
 \times\left((m_h^2-M_j^2)^2-4M_j^2|\vec{p}_{h_T}|^2\sin^2\phi_N\right)\bigg{]}^{1/2}\bigg{\}}
\end{multline}

\bibliographystyle{epjc}
\bibliography{displaced}

\end{document}